\documentclass[11pt,a4paper]{article}

\usepackage[T1]{fontenc}
\usepackage[utf8]{inputenc}
\usepackage{mathptmx}                 
\usepackage[scaled=0.92]{helvet}      
\usepackage{microtype}
\usepackage[margin=1in,headheight=14pt]{geometry}
\usepackage{setspace}
\usepackage{amsmath,amssymb,mathtools}
\usepackage{bm}

\usepackage{graphicx}
\usepackage{booktabs,tabularx,multirow,makecell,array}
\usepackage[font=small,labelfont=bf,skip=6pt]{caption}
\usepackage{subcaption}
\usepackage{enumitem}
\usepackage{float}
\usepackage{algorithm}
\usepackage{algpseudocode}
\graphicspath{{figures/}}

\usepackage[table]{xcolor}
\definecolor{tdblue}{HTML}{000000}
\definecolor{tdboxbg}{HTML}{EEF3FA}
\definecolor{tdboxframe}{HTML}{C9D6EA}
\definecolor{tdrule}{HTML}{2F5597}
\definecolor{tdgray}{HTML}{555555}
\usepackage[most]{tcolorbox}

\usepackage{titlesec}
\titleformat{\section}{\sffamily\Large\bfseries\color{tdblue}}{\thesection}{0.7em}{}
\titleformat{\subsection}{\sffamily\large\bfseries\color{tdblue}}{\thesubsection}{0.6em}{}
\titleformat{\subsubsection}{\sffamily\normalsize\bfseries}{\thesubsubsection}{0.5em}{}
\titlespacing*{\section}{0pt}{14pt plus 3pt minus 2pt}{6pt plus 2pt}
\titlespacing*{\subsection}{0pt}{10pt plus 2pt minus 2pt}{4pt plus 1pt}

\usepackage{fancyhdr}
\usepackage[numbers,square,sort&compress]{natbib}
\usepackage[colorlinks=true,linkcolor=tdblue,citecolor=tdblue,urlcolor=tdblue]{hyperref}
\hypersetup{pdfauthor={Siddhant Saxena},pdftitle={Learning Design Taste through Verifiable Synthetic Environments and Agentic Reinforcement Learning}}
\usepackage[capitalise,nameinlink]{cleveref}
\crefname{equation}{Eq.}{Eqs.}
\Crefname{equation}{Eq.}{Eqs.}
\crefname{algorithm}{Algorithm}{Algorithms}
\crefname{appendix}{Appendix}{Appendices}

\newcommand{\ind}{\mathbf{1}}                       
\newcommand{\Madm}{\mathcal{M}^{+}}                 
\newcommand{\Sfunc}{S_{\mathrm{func}}}
\newcommand{\Sdesign}{S_{\mathrm{design}}}
\newcommand{\Srubric}{S_{\mathrm{rubric}}}
\newcommand{\Spref}{S_{\mathrm{pref}}}
\newcommand{\code}[1]{\texttt{\small #1}}
\newcommand{\redgreen}{RED$\rightarrow$GREEN}
\newcommand{\best}[1]{\textbf{#1}}
\DeclareMathOperator{\clip}{clip}
\DeclareMathOperator{\sgn}{sgn}
\DeclareMathOperator{\slope}{slope}
\DeclareMathOperator{\acc}{acc}
\DeclareMathOperator{\mean}{mean}
\DeclareMathOperator{\std}{std}

\newcommand{\tablefont}{\small}
\newcolumntype{L}{>{\raggedright\arraybackslash}X}

\newtcolorbox{abstractbox}{
  enhanced, breakable,
  colback=tdboxbg, colframe=tdboxframe, boxrule=0.6pt, arc=6pt,
  left=12pt, right=12pt, top=8pt, bottom=10pt,
  fonttitle=\sffamily\bfseries\large, coltitle=tdblue,
  title=Abstract, center title,
  attach title to upper, after title={\par\smallskip},
}

\begin{document}

\thispagestyle{plain}
\begin{center}
  {\color{tdrule}\rule{\linewidth}{1.2pt}}\\[10pt]
  {\sffamily\bfseries\LARGE Learning Design Taste through Verifiable Synthetic\\[2pt]
   Environments and Agentic Reinforcement Learning\par}
  \vspace{6pt}
  {\color{tdrule}\rule{\linewidth}{0.6pt}}\\[10pt]
  {\large Sujay choubey \quad Siddhant Saxena\quad Sarthak Kapil \quad Suhas Motwani\quad \par}
  \vspace{2pt}
  {\small\ttfamily siddhant@basethesis.com\par}
  \vspace{4pt}
  {\normalsize October 2026\par}
\end{center}
\vspace{4pt}

\begin{abstractbox}
\noindent
Code-generating language models produce user interfaces that build and run, but the pages are often weak in hierarchy, spacing, typography and composition. We study whether a small code model can learn design taste through reinforcement learning in verifiable synthetic environments. The method has three components. First, a \textbf{layered verification framework} splits design quality into hard gates, render-grounded metrics, rubric judges, within-group relative preference and a blinded expert panel. A signal enters the reward only if it passes an admission protocol that tests agreement with experts, causal sensitivity, resistance to hack traps and stability. Second, following SCALER~\citep{xu2026scaler}, \textbf{synthetic design environments} are parameterised generators with measured difficulty ladders: base pages are written once by an agentic coder, and instances are produced by syntax-tree perturbation operators and constraint samplers, scheduled by an adaptive curriculum with a quarantine rule against reward over-optimisation. Third, a 4B code model is trained with \textbf{agentic reinforcement learning} over multi-turn episodes in which it edits and builds a repository and reads a textual report of its own render, while hidden graders compute the reward. The admission protocol admits 7 of 16 render signals and 6 of 10 rubric dimensions. On TD-Bench, agentic training raises the Taste Index from 0.630 to 0.795 in-family and from 0.598 to 0.700 on held-out families, and wins 0.600 of blinded expert comparisons against the base model and 0.569 against a single-turn control. Most of the advantage depends on the inference-time loop, and cross-brief distinctiveness falls slightly below that of the base model, the only success criterion not met.

\vspace{6pt}

\end{abstractbox}

\section{Introduction}
\label{sec:intro}

Language models now generate front-end code that builds and runs, yet the resulting interfaces are often weak in hierarchy, spacing, typography and composition. In Design2Code, generated pages were judged acceptable replacements for the reference in about half of cases, with layout the weakest dimension~\citep{si2025design2code}, and DesignBench reports the same deficit across frameworks~\citep{xiao2025designbench}. Design quality has no single correct answer, but it is not arbitrary: experienced designers agree on many failures, such as weak hierarchy, cramped spacing, low contrast and inconsistent components~\citep{td2026report}.

A common approach is a single-turn pipeline of supervised fine-tuning, preference optimisation and a short GRPO stage over a static corpus of generated pages~\citep{td2026pilot}. This has three limits. Static data loses its learning signal as the policy improves, as SCALER shows for reasoning tasks~\citep{xu2026scaler}, and a corpus written by one coder is bounded by that coder's taste. The webapp-synth substrate synthesises deterministically gradeable front-end tasks reliably, but its reward measures interaction rules rather than design, and its checks run in jsdom, which computes no layout~\citep{td2026webappsynth}. Finally, the strongest recent results on interface quality come from systems that let a model observe and revise its rendered output~\citep{relook2026,yang2025ui2coden,xiao2026aescoder,visionrefine2026}.

We address these limits with three components. Our contributions are:
\begin{enumerate}[label=(\roman*),leftmargin=2.2em,itemsep=1pt,topsep=2pt]
  \item a \textbf{five-layer verification framework} for design taste with an \textbf{admission protocol} that decides which signals may enter the reward (\cref{sec:verification});
  \item \textbf{synthetic design environments} adapted from SCALER, with a two-tier generation pipeline, seven families, perturbation operators and difficulty ladders calibrated from measured pass rates, scheduled by an adaptive curriculum with a \textbf{quarantine rule} (\cref{sec:environments});
  \item an \textbf{agentic training formulation} with visible development checks and hidden graders (\cref{sec:arl});
  \item \textbf{TD-Bench}, an evaluation with held-out families and blinded expert comparisons, and experiments against a single-turn control and a prompt-optimised agent (\cref{sec:setup,sec:results}).
\end{enumerate}

\section{Related Work}
\label{sec:related}

\paragraph{Evaluating and rewarding interface design.} Design2Code measures fidelity to a reference screenshot~\citep{si2025design2code}. ArtifactsBench scores rendered artefacts with a checklist-guided multimodal judge and reports over 90\% pairwise agreement with experts~\citep{artifactsbench2025}, UI-Bench fits TrueSkill ratings to 4,075 blinded expert comparisons~\citep{mateega2025uibench}, and UIClip is a pretrained quality scorer~\citep{wu2024uiclip}. For non-verifiable domains, Rubrics as Rewards uses instance-specific weighted rubrics as the GRPO reward~\citep{gunjal2025rar}. Model judges show position and verbosity biases~\citep{zheng2023judging}, and optimising any learned proxy eventually degrades true quality~\citep{gao2023scaling}. Relative formulations, such as pairwise visual comparison in UI2Code$^N$~\citep{yang2025ui2coden} and in-group win rates in Pref-GRPO~\citep{wang2025prefgrpo}, reduce calibration noise.

\paragraph{Render-in-the-loop generation.} AesCoder trains a 4B model with GRPO on execution and aesthetic rewards~\citep{xiao2026aescoder}. ReLook gives zero reward to pages that fail to render and accepts a revision only if a critic score improves~\citep{relook2026}. Vision-guided refinement improves frontend output by up to 17.8\% over three critique cycles~\citep{visionrefine2026}, and 1D-Bench defines round-level rewards for agentic interface editing~\citep{onedbench2026}.

\paragraph{Environments, curricula and agentic RL.} Reasoning Gym and RLVE provide procedurally generated environments with tunable difficulty~\citep{stojanovski2025reasoninggym,zeng2025rlve}. SCALER converts problems into environments with explicit scale parameters, validates their generators, adjusts difficulty online towards a target accuracy and retires stagnant environments~\citep{xu2026scaler}. GRPO removes the value network by normalising rewards within a group~\citep{shao2024deepseekmath}, SkyRL-Agent scales multi-turn rollouts asynchronously~\citep{skyrlagent2025}, and GEPA shows that prompt optimisation is a strong and inexpensive baseline~\citep{agrawal2026gepa}.

\paragraph{Learning taste.} Reinforcement Learning from Community Feedback trains a judge of research impact and uses in-group win rates as the reward for an ideation policy~\citep{tong2026scitaste}. TASTE finds that designer agreement is higher for rule-based criteria, such as typography, than for colour harmony~\citep{zhu2026taste}, and Taste-Bench measures agent taste as choosing the better direction at decision forks~\citep{pan2026tasteful}. Like these studies, we validate the reward against human judgement and use relative comparison where absolute scores are unreliable. Unlike them, we verify taste in layers, with deterministic render metrics carrying the parts on which experts agree.

\section{Problem Formulation and Overview}
\label{sec:formulation}

A task instance is a triple $x = (q, s_0, g)$: a natural-language instruction $q$ (a symptom description, a holistic request or a structured brief), an initial repository state $s_0$ (a complete page, a broken page or an empty scaffold) and a hidden grader configuration $g$. A policy $\pi_\theta$ interacts with a sandbox for at most $K$ turns, emitting tool calls or a submission and receiving observations. The production bundle of the submitted state is scored by graders the policy cannot observe, giving a terminal reward $R(\xi) \in [0, 1]$ for episode $\xi$.

Following RLVE and SCALER~\citep{xu2026scaler,zeng2025rlve}, an environment is a parameterised problem space with an explicit difficulty ladder,
\begin{equation}
  \mathcal{E}_e = \left( I_e, P_e, R_e, D_e \right), \qquad x = P_e(\omega, \ell), \quad \ell \in \{0, \ldots, D_e\},
  \label{eq:env}
\end{equation}
where $I_e$ is the instruction template and tool set, $P_e$ a generator that maps a seed $\omega$ and level $\ell$ to an instance, $R_e$ the verifier and $D_e$ the maximum level. Training maximises
\begin{equation}
  \max_\theta \; \mathbb{E}_{e \sim W_t,\, \ell \sim q_t^e} \, \mathbb{E}_{x \sim P_e(\omega, \ell)} \, \mathbb{E}_{\xi \sim \pi_\theta(\cdot \mid x)} \left[ R_e(\xi) \right],
  \label{eq:objective}
\end{equation}
where the active environment set $W_t$ and level distribution $q_t^e$ are adapted during training (\cref{sec:curriculum}). We do not assume a single verifier of design quality. We use signals $V_0, \ldots, V_4$ of increasing fidelity and decreasing reliability, and only candidates in the admitted set $\Madm \subseteq \mathcal{M}$, validated against the expert layer, contribute to $R$.

\begin{figure}[t]
  \centering
  \includegraphics[width=\linewidth]{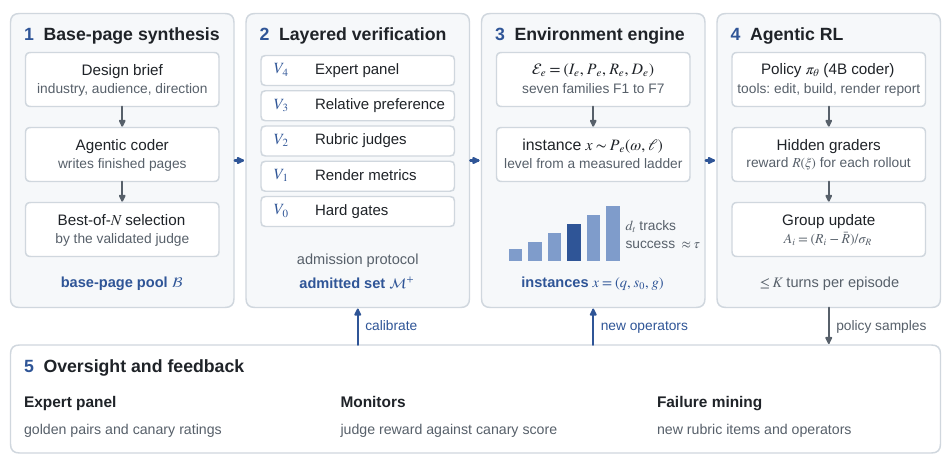}
  \caption{Overview. (1) Agentic generation of base pages with best-of-$N$ selection. (2) The five verification layers and the admission protocol that determines $\Madm$. (3) Environments that emit instances at a controlled level of a measured difficulty ladder. (4) Multi-turn agentic RL with hidden graders and group-relative advantages. (5) Oversight calibrates admission against experts, quarantines environments whose judge reward diverges from the canary, and converts failures into new rubric items and operators.}
  \label{fig:overview}
\end{figure}

\Cref{fig:overview} shows the system. The substrate is the webapp-synth repository, which provides a seed taxonomy, deterministic checks, a \redgreen{} admission gate and a sandboxed agent harness~\citep{td2026webappsynth}. Component A decides which signals measure improvement, Component B turns them into environments with difficulty ladders, and Component C trains the policy. Policy samples and failures return to Component A as expert labels, rubric items and perturbation operators.

\section{Component A: Layered Verification}
\label{sec:verification}

\begin{figure}[t]
  \centering
  \includegraphics[width=\linewidth]{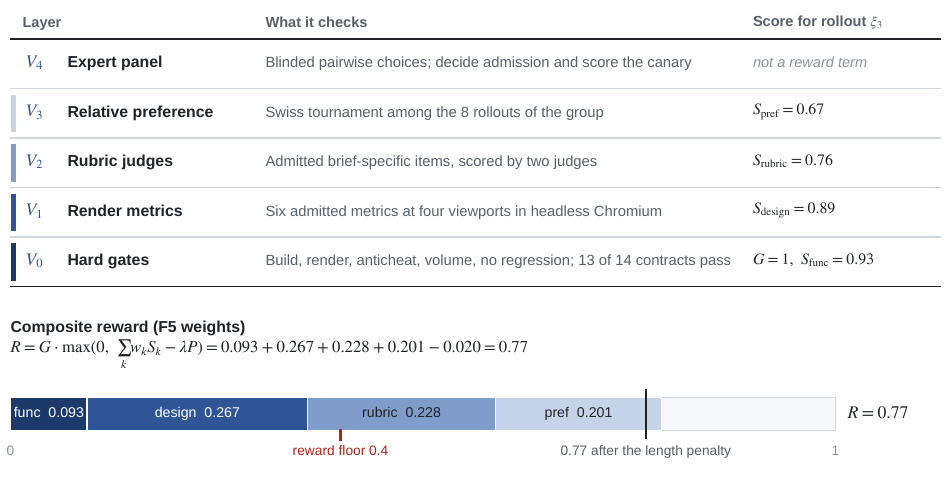}
  \caption{Layered verification for one rollout of a brief-to-page (F5) instance. The gates pass and 13 of 14 contract tests succeed ($\Sfunc = 0.93$); six render metrics give $\Sdesign = 0.89$; admitted rubric items give $\Srubric = 0.76$; an in-group tournament gives $\Spref = 0.67$. The expert layer decides admission and scores the canary but is not a reward term. With F5 weights, \cref{eq:reward} gives $R = 0.77$, above the 0.4 floor.}
  \label{fig:verify}
\end{figure}

Some parts of design quality follow explicit principles and can be computed from a rendered page, such as text contrast, adherence to a spacing scale and the absence of horizontal overflow on narrow viewports. Others need judgement but can be split into checkable questions, and some appear only under comparison. We therefore verify in five layers (\cref{fig:verify}). $V_0$ \textbf{hard gates} (build, render without errors, anticheat, page volume, functional contracts) are binary and multiply the reward. $V_1$ \textbf{render metrics}, $V_2$ \textbf{rubric judges} and $V_3$ \textbf{relative preference} contribute graded terms. $V_4$, a blinded \textbf{expert panel}, never enters the reward but decides which lower-layer signals may. Rendering uses headless Chromium on the production bundle, since $V_1$ and above need computed layout.

\paragraph{Render metrics ($V_1$).} Each metric $s_m$ maps the render capture $c_v$ at viewport $v$ (DOM, computed styles, box model and accessibility tree) to $[0, 1]$ and is averaged over viewports $\mathcal{V} = \{375, 768, 1280, 1440\}$\,px and interaction states. The library covers text contrast (WCAG AA~\citep{wcag22}), spacing-scale conformity, proximity, type-scale coherence, hierarchy integrity, primary-action salience, line length, alignment, palette discipline, component consistency, responsive integrity, accessibility, performance and token hygiene. Every metric reads computed styles rather than class names, and each has a recorded exploit and counterweight; for example, collapsing all spacing to one value raises scale conformity but lowers the proximity metric. Two further signals target generic output: a \textbf{default-tell} detector for common defaults, and the \textbf{slop distance}, the Mahalanobis distance of a page's style signature from the centroid of base-model pages. Neither measures quality, so both are used only in the art-direction family.

\paragraph{Rubric judges and preference ($V_2$, $V_3$).} Following RaR~\citep{gunjal2025rar}, each brief receives a weighted rubric over ten dimensions (hierarchy, layout, typography, colour, responsiveness, accessibility, copy, conversion, direction, craft) with negative-weight pitfall items. A \emph{bound} item is scored by its $V_1$ metric, so judges never re-estimate a measurable quantity; a \emph{judged} item is scored by the ensemble only if its template is in $\Madm$. The ensemble has a code-and-metrics judge and a screenshot judge from model families other than the policy's, and a third judge is withheld for evaluation. Comments and unrendered strings are stripped before judging, and pairs are judged in both orders~\citep{zheng2023judging}. Within a GRPO group of $n = 8$, a Swiss schedule of 12 comparisons replaces the 28 of a round robin, and $\Spref(\xi_i)$ is the win rate of rollout $i$ over its scheduled comparisons.

\paragraph{Expert panel ($V_4$).} Three to five senior designers give blinded forced-choice judgements, item scores and short rationales, with catch trials and repeats. Reliability is reported as Krippendorff's $\alpha$~\citep{krippendorff2018content}. Estimating an agreement rate within $\pm 0.05$ at 95\% confidence needs $N = 1.96^2 \times 0.25 / 0.05^2 \approx 385$ pairs per dimension.

\subsection{Admission protocol}
\label{sec:admission}
A candidate $m \in \mathcal{M}$ enters $\Madm$ only if it passes four tests. Let $\mathcal{P}_m$ be the golden pairs $(a, b, y)$, with expert preference $y \in \{-1, +1\}$, on which the metric difference exceeds its noise floor. \textbf{Agreement} is
\begin{equation}
  A_m = \frac{1}{|\mathcal{P}_m|} \sum_{(a, b, y) \in \mathcal{P}_m} \ind\!\left[ \sgn\!\left( s_m(a) - s_m(b) \right) = y \right] \geq 0.85.
  \label{eq:agreement}
\end{equation}
\textbf{Causal sensitivity} requires that on at least 90\% of single-principle minimal pairs the score drops after a perturbation and recovers after repair. \textbf{Hack-trap resistance} requires that the metric prefers a page built to game it over an expert-preferred page on at most 20\% of trap pairs. \textbf{Stability} requires at least 0.90 self-agreement across render seeds, viewport jitter and repeated judge calls. Failing items are kept for evaluation only, and admitted items are re-tested whenever the golden set or the judges change.

\subsection{Composite reward}
\label{sec:reward}
For a finished episode,
\begin{equation}
  R(\xi) = G(\xi) \cdot \max\!\Big( 0,\; \sum_{k} w_k S_k(\xi) - \lambda_{\mathrm{len}} P_{\mathrm{len}}(\xi) - \lambda_{\mathrm{turn}} P_{\mathrm{turn}}(\xi) \Big), \qquad k \in \{\mathrm{func}, \mathrm{design}, \mathrm{rubric}, \mathrm{pref}\},
  \label{eq:reward}
\end{equation}
with weights summing to one. The gate $G$ is the product of build, render, anticheat, volume and \textbf{no-regression} checks; the last gives zero reward if a repair degrades any other admitted metric of the base page, so deleting a defective section is never the cheapest fix. $\Sfunc$ is the vitest pass fraction~\citep{td2026webappsynth}, $\Sdesign$ the weighted mean of admitted metrics (restricted in repair families to the metrics affected by the planted defects), $\Srubric$ the score over admitted judged items, and $P_{\mathrm{len}}$, $P_{\mathrm{turn}}$ penalise page inflation and extra turns. Two rules from webapp-synth apply to every family: integrity is guarded by gates and never by the graded score, and the unmodified starting state must score at most 0.4 (the \emph{reward floor}). Instructions describe symptoms, never the fix.

\section{Component B: Synthetic Design Environments}
\label{sec:environments}

SCALER treats an environment, a program that yields unlimited verifiable instances of one problem type at a chosen difficulty, as the unit of training data~\citep{xu2026scaler}. In webapp-synth, by contrast, each instance needs a 20 to 30 minute agentic generation session and carries an unmeasured difficulty label~\citep{td2026webappsynth}. We keep the expensive generation step but run it once per base page, and generate instances with inexpensive programs.

\begin{figure}[t]
  \centering
  \includegraphics[width=\linewidth]{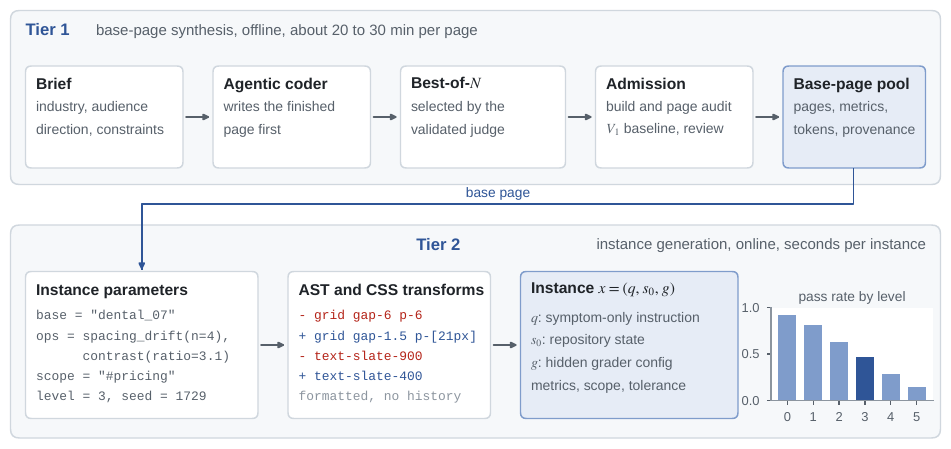}
  \caption{Two-tier environment synthesis. Tier 1 (offline) produces judge-selected base pages from briefs. Tier 2 (online) samples a base page and instance parameters (seed $\omega$, level $\ell$, operators), applies the transforms and emits an instance $x = (q, s_0, g)$. The inset gives the measured pass rate of the base policy at each level; the worked instance is at level 3.}
  \label{fig:synth}
\end{figure}

\paragraph{Two-tier generation.} \textbf{Tier 1} runs offline: agentic coding sessions write long-form pages from sampled briefs, the validated judge selects one of several candidates per brief, and selected pages that pass the build, the page audit, a $V_1$ baseline and a human spot check enter a \textbf{base-page pool}. \textbf{Tier 2} runs online: given a base page and a seed, a program samples perturbation operators and severities, applies them as syntax-tree and stylesheet edits, normalises formatting so the diff does not reveal edit locations, writes a symptom-only instruction and emits the hidden grader configuration in seconds (\cref{fig:synth}). Generation families use a constraint sampler instead. This split matches the cost asymmetry: a pool of several hundred base pages is affordable, tens of thousands of agentic sessions are not.

\paragraph{Families and operators.} Seven families target complementary aspects of design (\cref{tab:families}). Page-editing families (F1 to F4, F7) have a well-defined do-nothing floor and mostly deterministic rewards; open-ended families (F5, F6) rely more on judges and relative preference. Fourteen perturbation operators, such as spacing drift, contrast degradation, hierarchy scramble, action dilution and breakpoint removal, each break one principle and are each linked to the metric that detects it. Hidden graders test the violated invariant rather than restoration of the original source, so any correct repair receives full credit. F7 scores improvement over the starting page and accepts a round's edits only if the hidden score rises~\citep{relook2026}.

\begin{table}[t]
  \centering
  \caption{Environment families, their difficulty parameters and initial reward weights $w_k$ (\cref{eq:reward}). The weights keep every family's reward floor at or below 0.4.}
  \label{tab:families}
  \tablefont
  \begin{tabularx}{\linewidth}{@{}p{2.5cm}Lp{3.6cm}cccc@{}}
    \toprule
    \textbf{Family} & \textbf{Task} & \textbf{Difficulty parameters} & $\boldsymbol{\Sfunc}$ & $\boldsymbol{\Sdesign}$ & $\boldsymbol{\Srubric}$ & $\boldsymbol{\Spref}$ \\
    \midrule
    \textbf{F1} Interaction repair & Repair a broken interaction rule on a long-form page & Coupled seeds, localisation depth & 1.0 & & & \\
    \textbf{F2} Design polish & Locate and repair planted design defects & Number, subtlety and coupling of defects & & 0.8 & & 0.2 \\
    \textbf{F3} System adherence & Add a component using only given tokens and specifications & Token-set size, component novelty & 0.2 & 0.6 & 0.2 & \\
    \textbf{F4} Responsive adaptation & Make a desktop-only page correct at every viewport & Affected sections, layout complexity & 0.2 & 0.8 & & \\
    \textbf{F5} Brief to page & Build a page from a brief with explicit constraints & Constraint count, conflicting requirements & 0.1 & 0.3 & 0.3 & 0.3 \\
    \textbf{F6} Art direction & Build a page that commits to a named aesthetic direction & Specificity of direction & & 0.2 & 0.4 & 0.4 \\
    \textbf{F7} Critique revision & Improve a page over several rounds from a critique & Critique specificity, rounds & & 0.4 & & 0.6 \\
    \bottomrule
  \end{tabularx}
\end{table}

\paragraph{Generator validation and difficulty.} Because Tier 2 generators are partly written by an agent, each must pass SCALER-style checks~\citep{xu2026scaler,fu2025klear} before entering the pool: the \redgreen{} gate (the instance fails a hidden check, the base page passes all), breadth (two independent correct repairs both pass), depth (instances at one level differ in target elements), the reward floor, volume bounds, absence of leakage, and severity monotonicity. Design has no single scale parameter, so levels are defined empirically: the base policy is run on a grid of operator subsets and severities, and a monotone fit maps grid cells to levels $0, \ldots, D_e$ with roughly uniform steps in pass rate.

\subsection{Adaptive curriculum}
\label{sec:curriculum}
\begin{figure}[t]
  \centering
  \includegraphics[width=\linewidth]{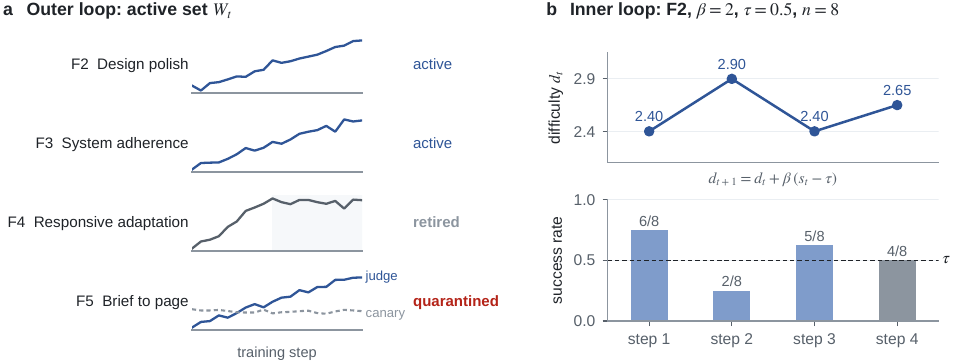}
  \caption{Adaptive curriculum. (a) Outer loop: each line is the difficulty of one environment over training. F2 and F3 are still rising and stay active; F4 stagnates and is retired temporarily; F5's judge reward rises while its canary score stays flat, so it is quarantined. (b) Inner loop for one F2 environment ($\beta = 2$, $\tau = 0.5$, $n = 8$): the success rate of each group moves the difficulty by \cref{eq:controller}.}
  \label{fig:curriculum}
\end{figure}

An episode is a success if it passes every gate and its reward exceeds a family threshold $\theta_e$. With success rate $\acc_t^e$ over the $k \times n$ rollouts of environment $e$ at step $t$, the difficulty is updated as
\begin{equation}
  d_{t+1}^e = \clip\!\left( d_t^e + \beta \left( \acc_t^e - \tau \right),\, 0,\, D_e \right),
  \label{eq:controller}
\end{equation}
and the $k$ instances of a step are split between the two nearest integer levels so that their mean approximates $d$~\citep{xu2026scaler}. A target $\tau \approx 0.5$ maximises the chance that a group contains both successes and failures, and hence a non-zero advantage. The outer loop keeps an active set $W_t$ of fixed size and, as in SCALER, retires an environment temporarily when its difficulty slope over 10 steps is not positive, its success rate is zero for 5 steps, or its difficulty stays at $D_e$ for 5 steps. We add one rule for learned rewards: an environment is \textbf{quarantined} for human review when its judge-derived reward rises while its canary score (expert or held-out judge) stays flat over $K_{\mathrm{hack}}$ steps,
\begin{equation}
  Q_e = \ind\!\left[ \slope\!\left( \bar{R}_{\mathrm{judge}}^e \right) > 0 \wedge \slope\!\left( \bar{S}_{\mathrm{canary}}^e \right) \leq 0 \right],
  \label{eq:quarantine}
\end{equation}
since rising proxy reward with flat true quality is the typical sign of over-optimisation~\citep{gao2023scaling}. \Cref{fig:curriculum} shows both loops.

\paragraph{Held-out environments.} One complete family (F4), two operator types, one industry and a set of designer-authored briefs are excluded from training. A policy that has learned to invert specific operators, rather than to perceive design problems, will show a large gap on them.

\section{Component C: Agentic Reinforcement Learning}
\label{sec:arl}

A single-turn policy must emit a finished page without seeing it, whereas design practice is iterative, and models benefit from observing and revising their output~\citep{visionrefine2026,relook2026}. Multi-turn episodes also make the repair families possible, since they are operations on a stateful repository. The costs are sandbox time, harder credit assignment and malformed tool calls from a small model, so we keep a single-turn control and a prompt-optimised agent as baselines.

\begin{figure}[t]
  \centering
  \includegraphics[width=\linewidth]{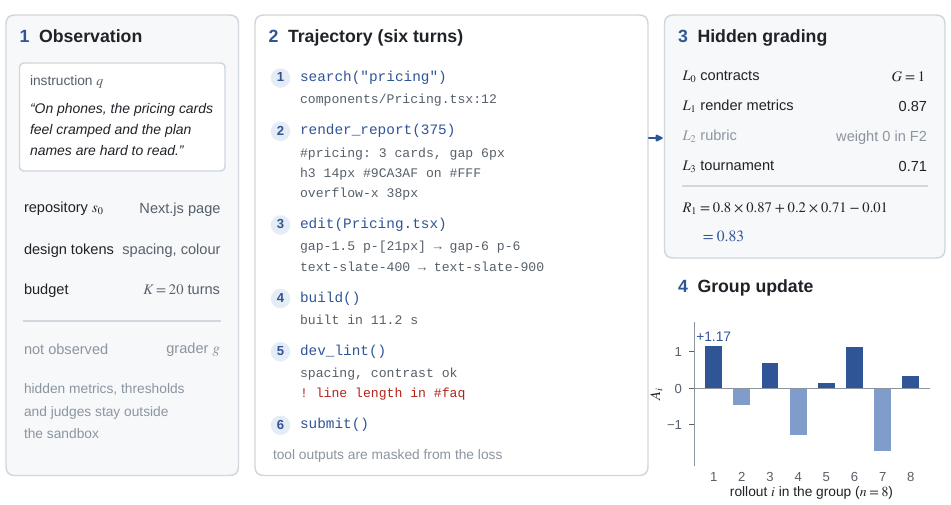}
  \caption{An agentic episode. (1) The policy sees a symptom-only instruction, the repository, design tokens and its budget, but not the grader. (2) It locates the component, inspects the render report at 375\,px, edits spacing and colour classes, rebuilds, runs the linter and submits. (3) Hidden graders score the bundle; with F2 weights, $R_1 = 0.83$. (4) Rewards in the group become advantages (\cref{eq:advantage}) for the policy update.}
  \label{fig:episode}
\end{figure}

\paragraph{Episodes and tools.} The observation holds the instruction, the repository, any design tokens and a budget of 20 turns (\cref{fig:episode}). The policy can read, search and edit files, build, call \code{render\_report(v)} for a textual outline of the render at viewport $v$ (headings with computed sizes, per-section layout, overflowing or clipped elements), run public functional tests, and call \code{dev\_lint}. The render report describes what is present, not whether it is good. The key design decision is how much of the verifier to expose. Following the convention of public and hidden tests, \code{dev\_lint} covers a rotating subset of metrics at relaxed thresholds, while hidden graders cover all admitted metrics at brief-specific thresholds together with the judges. We monitor the visible--hidden gap $\Gamma_t = \bar{S}_{\mathrm{vis},t} - \bar{S}_{\mathrm{hid},t}$ on shared metrics; a rising $\Gamma_t$ shows that the policy is fitting the visible checks.

\paragraph{Objective.} For $n = 8$ rollouts of an instance, advantages are group-normalised~\citep{shao2024deepseekmath},
\begin{equation}
  A_i = \frac{R_i - \mean(R_1, \ldots, R_n)}{\std(R_1, \ldots, R_n) + \epsilon},
  \label{eq:advantage}
\end{equation}
and the policy maximises the clipped surrogate with a token mask that excludes tool outputs from the loss, plus a small KL penalty to a reference policy, kept because the reward includes learned judges. We use a single terminal reward per episode: it is the hardest to exploit, and potential-based shaping~\citep{ng1999policy} telescopes into an end-of-episode term under trajectory-level GRPO. The policy is a 4B code model (Qwen3.5-4B) with a short warm-up on tool-use traces replayed from the 48 existing brief--page pairs.

\paragraph{System.} Rendering and judging run outside the agent sandbox. The policy builds inside a prebaked sandbox image (up to 256 concurrent), which also runs the functional tests. The production bundle is sent to a separate render and reward service with a pool of headless browsers, the metric library and the judge ensemble, which caches results by bundle hash. This keeps the sandbox small, lets renders be batched and cached across rollouts, and makes the hidden graders unreachable from the policy. Monitors read the metric stream and can quarantine an environment through the curriculum controller. Cost is accounted in sandbox-minutes per solved instance.

\section{Experimental Setup}
\label{sec:setup}

\paragraph{Hypotheses.} \textbf{H1}: a measurable share of expert preferences is predictable from render metrics, with the share varying by dimension. \textbf{H2}: adaptive environments keep the learning signal more informative than a static task set. \textbf{H3}: at matched compute, multi-turn agentic RL achieves a higher expert win rate than single-turn training, and part of the gain persists without tools. \textbf{H4}: the layered reward keeps the divergence between reward gain and expert gain small. \textbf{H5}: a code-and-metrics judge agrees with experts nearly as well as a screenshot judge, except on composition.

\paragraph{Experiments.} E1 applies the admission protocol to every candidate signal (H1). E2 measures the share of informative GRPO groups (H2). E3 compares training methods at matched compute (H3). E4 measures robustness diagnostics (H4). E5 compares the two judge types per dimension (H5). E6 removes tools at inference (H3). E7 measures distinctiveness across briefs.

\paragraph{Benchmark.} TD-Bench extends MiniLandingBench~\citep{td2026pilot} with four splits: held-out instances of training families (S1, in-family); held-out families, operators and industry (S2, held-out); tool-free single-shot generation from designer briefs; and an external slice from OpenDesign~\citep{xiao2026aescoder} and ArtifactsBench~\citep{artifactsbench2025}.

\paragraph{Systems.} All systems use the same 4B policy, harness, briefs and render service, are matched in sandbox-minutes and run with three seeds. \textbf{BASE} is the untrained model in the agentic harness. \textbf{SFT-C} is the single-turn control: SFT on the 48 long-form pages, DPO on golden pairs~\citep{rafailov2023dpo} and a short single-turn GRPO stage. \textbf{GEPA} is the base agent with a system prompt evolved by GEPA~\citep{agrawal2026gepa}, testing whether gains need training. \textbf{ARL} is our method.

\paragraph{Metrics.} The primary metrics are gate pass rate (P1), functional contract rate (P2), four composites of admitted and evaluation-only render metrics (legibility P3, system coherence P4, hierarchy and composition P5, render robustness P6), rubric score (P7), brief-constraint retention (P8), blinded expert win rate (P9) and distinctiveness (P10). The \textbf{Taste Index} (TI) is the unweighted mean of P3 to P8; expert win rate is kept separate as ground truth. Win rates carry Wilson 95\% intervals. A two-sided test at $\alpha = 0.05$ with power 0.8 needs about 194 comparisons to detect a win rate of 0.60 and 783 for 0.55, so we allocate 400 expert comparisons for ARL against BASE on S2, 800 against SFT-C, and 200 each against GEPA and in-family.

\paragraph{Success criteria.} Fixed before the experiments, on held-out splits: an expert win rate of at least 0.60 against BASE and 0.55 against SFT-C; no loss in gate pass rate or cross-brief diversity relative to BASE; reward--expert divergence at most 0.05; held-out gains at least half of in-family gains; and no unresolved quarantine events.

\section{Results}
\label{sec:results}

\subsection{What is verifiable (E1, E5)}
\label{sec:res-verifiable}
\begin{figure}[t]
  \centering
  \includegraphics[width=0.96\linewidth]{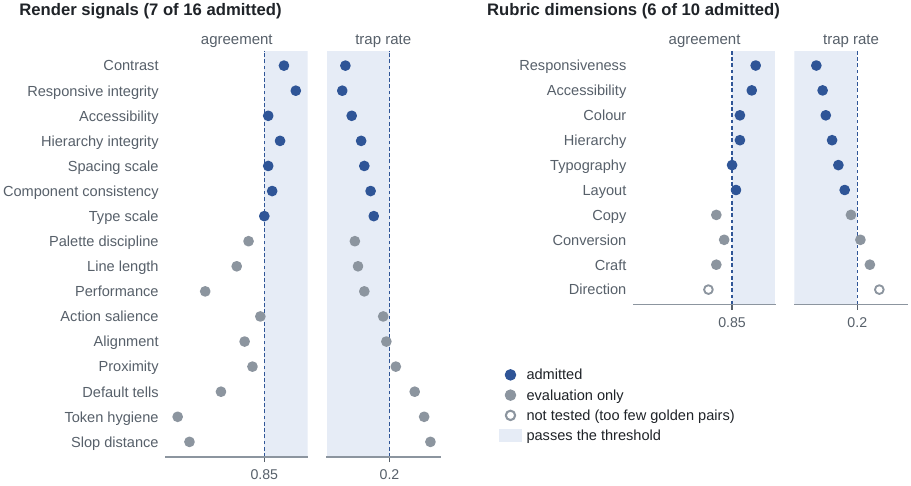}
  \caption{Admission outcomes (E1). Each row is a candidate signal, with its agreement with expert pairs (left) and hack-trap preference rate (right). Shaded regions satisfy the thresholds (agreement at least 0.85, trap rate at most 0.20). Blue points are admitted, grey points are evaluation-only, and the open point was not tested.}
  \label{fig:res-admission}
\end{figure}

The admission protocol admits 7 of the 16 render signals and 6 of the 10 rubric dimensions (\cref{fig:res-admission}). The admitted signals (contrast, spacing scale, type scale, component consistency, hierarchy, responsiveness and accessibility) are tied to explicit standards or to defects that designers name consistently. Line length, palette discipline, action salience, proximity and alignment fall short on agreement or exceed the trap ceiling, and token hygiene fails because experts judge pixels rather than class names. Copy, conversion and craft remain evaluation-only, and direction could not be tested with only 240 golden pairs. Composition is the largest verification gap: of its four signals, only hierarchy integrity enters the reward. The judge comparison supports H5. The screenshot judge agrees with experts clearly better only on layout ($+0.14$) and direction ($+0.11$), the code-and-metrics judge is better on accessibility ($-0.07$), where roles and focus order are invisible in pixels, and on the other seven dimensions the two differ by at most 0.05.

\subsection{Agentic training against the controls (E3)}
\label{sec:res-main}
\begin{figure}[t]
  \centering
  \includegraphics[width=0.9\linewidth]{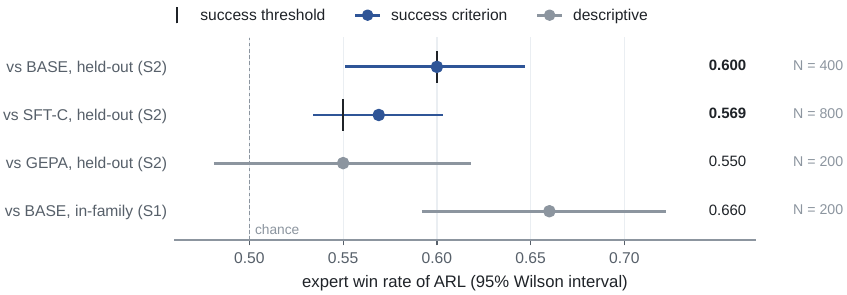}
  \caption{Expert win rates of ARL with 95\% Wilson intervals. Vertical ticks mark the success thresholds and the dashed line marks chance.}
  \label{fig:res-winrates}
\end{figure}

In-family, ARL raises the Taste Index from 0.630 to 0.795, against 0.689 for SFT-C and 0.690 for GEPA (\cref{tab:main}). The largest gains are on metrics carried by admitted signals and heavy family weights: gate pass rate (0.70 to 0.93), functional contracts (0.52 to 0.81) and legibility (0.74 to 0.88). The smallest are on hierarchy and composition (0.61 to 0.77) and rubric score (0.52 to 0.67), which depend on evaluation-only signals and noisy judged items. On the held-out split, ARL wins 0.600 of expert comparisons against BASE (Wilson interval 0.551 to 0.647, $N = 400$), exactly at the threshold, and 0.569 against SFT-C (0.534 to 0.603, $N = 800$), which meets the 0.55 criterion (\cref{fig:res-winrates}). The contrast with GEPA (0.55, $N = 200$) is reported descriptively.

\begin{table}[t]
  \centering
  \caption{Primary metrics on the in-family (S1) and held-out (S2) splits. The best value in each split is in bold. P9 is the expert win rate against BASE.}
  \label{tab:main}
  \tablefont
  \begin{tabular}{@{}l cccc c cccc@{}}
    \toprule
    & \multicolumn{4}{c}{\textbf{In-family (S1)}} && \multicolumn{4}{c}{\textbf{Held-out (S2)}} \\
    \cmidrule{2-5}\cmidrule{7-10}
    \textbf{Metric} & BASE & SFT-C & GEPA & ARL && BASE & SFT-C & GEPA & ARL \\
    \midrule
    Gate pass rate (P1)          & 0.70 & 0.79 & 0.78 & \best{0.93} && 0.68 & 0.73 & 0.74 & \best{0.82} \\
    Functional contracts (P2)    & 0.52 & 0.62 & 0.58 & \best{0.81} && 0.48 & 0.53 & 0.53 & \best{0.66} \\
    Legibility (P3)              & 0.74 & 0.80 & 0.79 & \best{0.88} && 0.73 & 0.76 & 0.77 & \best{0.82} \\
    System coherence (P4)        & 0.66 & 0.73 & 0.69 & \best{0.83} && 0.64 & 0.68 & 0.67 & \best{0.75} \\
    Hierarchy \& composition (P5) & 0.61 & 0.67 & 0.64 & \best{0.77} && 0.59 & 0.62 & 0.62 & \best{0.69} \\
    Render robustness (P6)       & 0.70 & 0.74 & 0.74 & \best{0.86} && 0.60 & 0.62 & 0.63 & \best{0.70} \\
    Rubric score (P7)            & 0.52 & 0.58 & 0.60 & \best{0.67} && 0.50 & 0.53 & 0.57 & \best{0.60} \\
    Constraint retention (P8)    & 0.55 & 0.61 & 0.69 & \best{0.75} && 0.53 & 0.56 & \best{0.65} & \best{0.65} \\
    Distinctiveness (P10)        & 0.48 & 0.44 & \best{0.55} & 0.45 && 0.47 & 0.45 & \best{0.53} & 0.45 \\
    \midrule
    Taste Index (TI)             & 0.630 & 0.689 & 0.690 & \best{0.795} && 0.598 & 0.631 & 0.650 & \best{0.700} \\
    Win rate vs BASE (P9)        & n/a & 0.559 & 0.560 & \best{0.659} && n/a & 0.532 & 0.551 & \best{0.601} \\
    \bottomrule
  \end{tabular}
\end{table}

\subsection{Generalisation and the role of the loop (E6)}
\label{sec:res-loop}
\begin{figure}[t]
  \centering
  \includegraphics[width=0.86\linewidth]{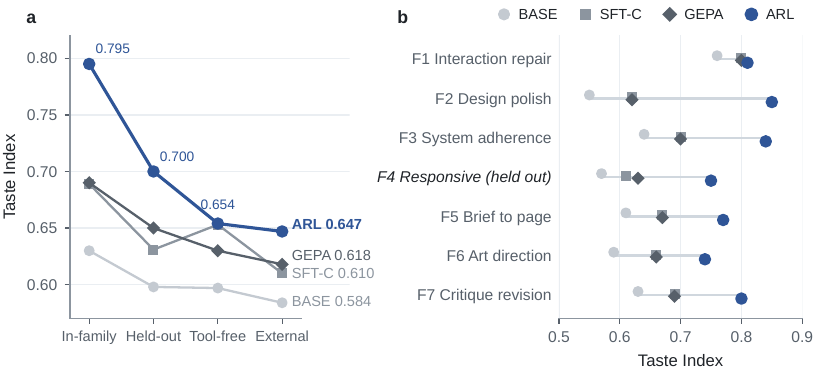}
  \caption{(a) Taste Index on the four TD-Bench splits. (b) Taste Index by environment family; F4 is held out.}
  \label{fig:res-general}
\end{figure}

ARL keeps 62\% of its in-family gain on held-out families, above the required half (\cref{fig:res-general}a). The advantage depends on the loop: with tools removed, ARL keeps only 35\% of its gain (TI 0.654) and matches SFT-C (0.653), which was trained single-turn. Part of the gain therefore persists, as H3 states, but agentic training does not give a better tool-free policy. On the external slice ARL remains ahead (0.647 against 0.610 to 0.618). By family (\cref{fig:res-general}b), ARL gains most where the reward is deterministic, design polish (0.55 to 0.85) and system adherence (0.64 to 0.84), and the held-out responsive family reaches 0.75.

\subsection{Robustness, curriculum and cost (E2, E4)}
\label{sec:res-robust}
\begin{figure}[t]
  \centering
  \includegraphics[width=\linewidth]{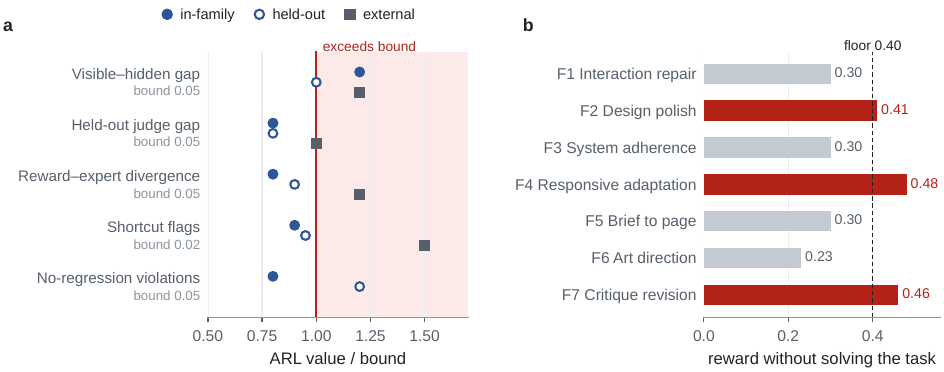}
  \caption{(a) ARL diagnostics divided by their bounds; points in the shaded region exceed the bound. (b) Reward earned without solving the task, per family; red bars exceed the 0.4 floor.}
  \label{fig:res-robust}
\end{figure}

The divergence between reward gain and expert gain stays within the 0.05 bound in-family and on held-out families (0.040 and 0.045) and exceeds it only on the external slice (0.06) (\cref{fig:res-robust}a). The visible--hidden gap reaches 0.06 in-family because \code{dev\_lint} is visible throughout training, and no-regression violations reach 0.06 under held-out operators. A separate check measures the reward a policy earns without solving the task (\cref{fig:res-robust}b). Three families exceed the 0.4 floor: in design polish (0.41) perturbed metrics remain partly high, in responsive adaptation (0.48) an untouched desktop page keeps its functional contracts, and in critique revision (0.46) an unchanged page can outrank worse revisions. Scoring $\Sdesign$ as the repaired fraction, restricting $\Sfunc$ in F4 to responsive contracts, and scoring F7 by improvement remove these violations.

The adaptive curriculum keeps 0.86 of GRPO groups informative, with both successes and failures, against 0.55 for the static-data GRPO stage of SFT-C, which supports H2 for the learning signal; its effect on expert win rate was not isolated. ARL is also the least costly system per solved instance, 6.9 sandbox-minutes against 14.7 for BASE, because its higher success rate outweighs its longer episodes. Brief-constraint retention rises from 0.55 to 0.75, but constraints that must survive an edit are lost twice as often as explicit ones, and conflicting requirements are flagged in only about half of briefs.

\subsection{Distinctiveness and external comparison (E7)}
\label{sec:res-distinct}
Distinctiveness is the only success criterion ARL does not meet (\cref{tab:criteria}). Slop distance and default tells measure departure from generic output rather than quality, so they cannot pass the agreement test, and direction has too few golden pairs. Without these counterweights, 80\% of the F6 reward comes from judge terms that favour the judges' central taste. On the held-out split, ARL's distinctiveness falls from 0.47 to 0.454, while GEPA, whose prompt asks for an explicit direction, rises to 0.532. On OpenDesign, ARL scores 78.4 against 75.0 for BASE, seventh of sixteen published systems and 3.5 points below AesCoder-4B, which is trained with a screenshot reward on similar data~\citep{xiao2026aescoder}.

\begin{table}[t]
  \centering
  \caption{Success criteria applied to the held-out (S2) results.}
  \label{tab:criteria}
  \tablefont
  \begin{tabular}{@{}llll@{}}
    \toprule
    \textbf{Criterion} & \textbf{Target} & \textbf{Result} & \textbf{Outcome} \\
    \midrule
    Expert win rate against BASE   & $\geq 0.60$ & 0.600 (0.551--0.647) & Pass, no margin \\
    Expert win rate against SFT-C  & $\geq 0.55$ & 0.569 (0.534--0.603) & Pass \\
    Gate pass rate                 & $\geq$ BASE (0.68) & 0.82 & Pass \\
    Cross-brief diversity          & $\geq$ BASE (0.47) & 0.454 & \textbf{Fail} \\
    Reward--expert divergence      & $\leq 0.05$ & 0.045 & Pass \\
    Held-out gain $\div$ in-family gain & $\geq 0.50$ & 0.62 & Pass \\
    Unresolved quarantine events   & 0 & 0 & Pass \\
    \midrule
    \textbf{Criteria met} & & & \textbf{6 of 7} \\
    \bottomrule
  \end{tabular}
\end{table}

\section{Discussion and Limitations}
\label{sec:discussion}

The verifiability map agrees with recent work on designer preference. TASTE reports higher designer agreement for typography and spatial accuracy than for colour harmony~\citep{zhu2026taste}, and in our results the signals tied to explicit rules pass admission while composition and direction do not. Taste-Bench finds that judgement errors concentrate where the deciding evidence appears late in a trajectory~\citep{pan2026tasteful}; the dependence of our gains on the render loop is consistent with this, since the loop lets the policy see the effect of its edits before submitting. The results point to four changes: admit distinctiveness terms through a direction-adherence test rather than expert agreement, which addresses the unmet criterion; rescore F2, F4 and F7 so every family respects the reward floor; rotate the \code{dev\_lint} subset per epoch to reduce $\Gamma$; and add a screenshot-judged layout item to close part of the composition gap.

\paragraph{Limitations.} The rubric and metric library may encode a single notion of taste; we keep label distributions where expert agreement is low and condition items on the requested direction. Base pages come from one agentic coder, so judges are drawn from other model families and one judge is held out. The scope is long-form landing pages in React and Tailwind; transfer to dashboards, multi-screen flows or native interfaces is not established. The primary win rate against BASE sits exactly at its threshold and should be read with its interval. The policy is text-only, only one size (4B) was trained, and neither the curriculum's effect on expert win rate nor scaling with the number of environments was measured.

\section{Conclusion}
\label{sec:conclusion}

We presented a method in which a small code model learns design taste by reinforcement learning in environments that make taste checkable in layers. Render metrics and rubric judges enter the reward only through an admission protocol validated against experts, environments are generators with measured difficulty scheduled by a curriculum that also guards against over-optimisation, and the policy learns through multi-turn episodes scored by graders it cannot observe. About half of the candidate signals can be verified, with composition the largest gap. Agentic training improves every primary metric except distinctiveness, transfers to held-out families, and meets the expert win-rate criteria against the base model and the single-turn control, although most of the gain depends on the inference-time loop. Admitting distinctiveness through direction adherence and closing the composition gap are the main directions for future work.

\bibliographystyle{unsrtnat}
\bibliography{references}

@article{xu2026scaler,
  title   = {{SCALER}: Synthetic Scalable Adaptive Learning Environment for Reasoning},
  author  = {Xu, C. and Xiao, C. and Peng, Z. and Wang, X. and Cao, Y.},
  journal = {arXiv preprint arXiv:2601.04809v5},
  year    = {2026}
}

@inproceedings{si2025design2code,
  title     = {{Design2Code}: Benchmarking Multimodal Code Generation for Automated Front-End Engineering},
  author    = {Si, C. and Zhang, Y. and Li, R. and Yang, Z. and Liu, R. and Yang, D.},
  booktitle = {Proceedings of the Conference of the Nations of the Americas Chapter of the Association for Computational Linguistics (NAACL)},
  pages     = {3956--3974},
  year      = {2025}
}

@article{xiao2025designbench,
  title   = {{DesignBench}: A Comprehensive Benchmark for {MLLM}-Based Front-End Code Generation},
  author  = {Xiao, J. and Wang, M. and Lam, M. H. and Wan, Y. and Liu, J. and Huo, Y. and Lyu, M. R.},
  journal = {arXiv preprint arXiv:2506.06251},
  year    = {2025}
}

@techreport{td2026report,
  title       = {Teaching {LLMs} {UI} Design Taste: Post-Training, {RL} Alignment, Agentic {RL}, Synthetic Environments, Design-Taste Evaluation and Meta-Harnesses},
  author      = {{Tasteful Designs Project}},
  institution = {Tasteful Designs Project},
  type        = {Internal research report},
  month       = sep,
  year        = {2026}
}

@techreport{td2026pilot,
  title       = {Developing Taste for Design in Models: Rubric-Grounded Golden Preferences and Reinforcement Post-Training of a Small {LLM} for Generating Better Design},
  author      = {{Tasteful Designs Project}},
  institution = {Tasteful Designs Project},
  type        = {Internal pilot design},
  month       = jun,
  year        = {2026}
}

@misc{td2026webappsynth,
  title        = {webapp-synth: Architecture and Operations Guide, and {README}},
  author       = {{Tasteful Designs Project}},
  howpublished = {Repository documentation, \url{https://github.com/mr-siddy/Tasteful-Designs}},
  year         = {2026}
}

@inproceedings{relook2026,
  title     = {{ReLook}: Vision-Grounded {RL} with a Multimodal {LLM} Critic for Agentic Web Coding},
  author    = {{ReLook authors}},
  booktitle = {Proceedings of the Annual Meeting of the Association for Computational Linguistics (ACL)},
  note      = {arXiv:2510.11498},
  year      = {2026}
}

@article{yang2025ui2coden,
  title   = {{UI2Code$^N$}: {UI}-to-Code Generation as Interactive Visual Optimization},
  author  = {Yang, Z. and Hong, W. and others},
  journal = {arXiv preprint arXiv:2511.08195},
  year    = {2025}
}

@inproceedings{xiao2026aescoder,
  title     = {Code Aesthetics with Agentic Reward Feedback},
  author    = {Xiao, B. and Jiang, L. and Huang, S. and others},
  booktitle = {International Conference on Learning Representations (ICLR)},
  note      = {arXiv:2510.23272},
  year      = {2026}
}

@article{visionrefine2026,
  title   = {Vision-Guided Iterative Refinement for Frontend Code Generation},
  author  = {{Vision-guided refinement authors}},
  journal = {arXiv preprint arXiv:2604.05839},
  year    = {2026}
}

@article{artifactsbench2025,
  title   = {{ArtifactsBench}: Bridging the Visual-Interactive Gap in {LLM} Code Generation Evaluation},
  author  = {{Tencent Hunyuan}},
  journal = {arXiv preprint arXiv:2507.04952},
  year    = {2025}
}

@article{mateega2025uibench,
  title   = {{UI-Bench}: A Benchmark for Evaluating Design Capabilities of {AI} Text-to-App Tools},
  author  = {Mateega, S. and others},
  journal = {arXiv preprint arXiv:2508.20410},
  year    = {2025}
}

@inproceedings{wu2024uiclip,
  title     = {{UIClip}: A Data-Driven Model for Assessing User Interface Design},
  author    = {Wu, J. and others},
  booktitle = {Proceedings of the ACM Symposium on User Interface Software and Technology (UIST)},
  year      = {2024}
}

@article{gunjal2025rar,
  title   = {Rubrics as Rewards: Reinforcement Learning beyond Verifiable Domains},
  author  = {Gunjal, A. and Wang, A. and Lau, E. and Nath, V. and He, Y. and Liu, B. and Hendryx, S.},
  journal = {arXiv preprint arXiv:2507.17746},
  year    = {2025}
}

@inproceedings{zheng2023judging,
  title     = {Judging {LLM}-as-a-Judge with {MT-Bench} and {Chatbot Arena}},
  author    = {Zheng, L. and Chiang, W.-L. and Sheng, Y. and Zhuang, S. and Wu, Z. and Zhuang, Y. and Lin, Z. and Li, Z. and Li, D. and Xing, E. P. and Zhang, H. and Gonzalez, J. E. and Stoica, I.},
  booktitle = {Advances in Neural Information Processing Systems (NeurIPS)},
  year      = {2023}
}

@inproceedings{gao2023scaling,
  title     = {Scaling Laws for Reward Model Overoptimization},
  author    = {Gao, L. and Schulman, J. and Hilton, J.},
  booktitle = {International Conference on Machine Learning (ICML)},
  note      = {arXiv:2210.10760},
  year      = {2023}
}

@article{wang2025prefgrpo,
  title   = {{Pref-GRPO}: Pairwise Preference Reward-Based {GRPO} for Stable Text-to-Image Reinforcement Learning},
  author  = {Wang, Y. and Li, Z. and Zang, Y. and Zhou, Y. and Bu, J. and Wang, C. and Lu, Q. and Jin, C. and Wang, J.},
  journal = {arXiv preprint arXiv:2508.20751},
  year    = {2025}
}

@article{onedbench2026,
  title   = {{1D-Bench}},
  author  = {{1D-Bench authors}},
  journal = {arXiv preprint arXiv:2602.18548},
  year    = {2026}
}

@article{stojanovski2025reasoninggym,
  title   = {Reasoning Gym: Reasoning Environments for Reinforcement Learning with Verifiable Rewards},
  author  = {Stojanovski, Z. and Stanley, O. and Sharratt, J. and Jones, R. and Adefioye, A. and Kaddour, J. and K{\"o}pf, A.},
  journal = {arXiv preprint arXiv:2505.24760},
  year    = {2025}
}

@article{zeng2025rlve,
  title   = {{RLVE}: Scaling Up Reinforcement Learning for Language Models with Adaptive Verifiable Environments},
  author  = {Zeng, Z. and Ivison, H. and Wang, Y. and Yuan, L. and Li, S. S. and Ye, Z. and Li, S. and He, J. and Zhou, R. and Chen, T. and others},
  journal = {arXiv preprint arXiv:2511.07317},
  year    = {2025}
}

@article{skyrlagent2025,
  title   = {{SkyRL-Agent}: Efficient {RL} Training for Multi-Turn {LLM} Agents},
  author  = {{SkyRL team}},
  journal = {arXiv preprint arXiv:2511.16108},
  year    = {2025}
}

@article{shao2024deepseekmath,
  title   = {{DeepSeekMath}: Pushing the Limits of Mathematical Reasoning in Open Language Models},
  author  = {Shao, Z. and Wang, P. and Zhu, Q. and Xu, R. and Song, J. and Zhang, M. and Li, Y. K. and Wu, Y. and Guo, D.},
  journal = {arXiv preprint arXiv:2402.03300},
  year    = {2024}
}

@inproceedings{agrawal2026gepa,
  title     = {{GEPA}: Reflective Prompt Evolution Can Outperform Reinforcement Learning},
  author    = {Agrawal, L. A. and others},
  booktitle = {International Conference on Learning Representations (ICLR)},
  note      = {arXiv:2507.19457},
  year      = {2026}
}

@misc{wcag22,
  title        = {Web Content Accessibility Guidelines ({WCAG}) 2.2},
  author       = {{World Wide Web Consortium}},
  howpublished = {W3C Recommendation},
  year         = {2023}
}

@book{krippendorff2018content,
  title     = {Content Analysis: An Introduction to Its Methodology},
  author    = {Krippendorff, K.},
  edition   = {4th},
  publisher = {SAGE Publications},
  year      = {2018}
}

@article{fu2025klear,
  title   = {{Klear-CodeTest}: Scalable Test Case Generation for Code Reinforcement Learning},
  author  = {Fu, J. and Yang, X. and Zhang, H. and Liu, Y. and Zhang, J. and Wang, Q. and Zhang, F. and Zhou, G.},
  journal = {arXiv preprint arXiv:2508.05710},
  year    = {2025}
}

@inproceedings{ng1999policy,
  title     = {Policy Invariance under Reward Transformations: Theory and Application to Reward Shaping},
  author    = {Ng, A. Y. and Harada, D. and Russell, S.},
  booktitle = {International Conference on Machine Learning (ICML)},
  year      = {1999}
}

@inproceedings{rafailov2023dpo,
  title     = {Direct Preference Optimization: Your Language Model Is Secretly a Reward Model},
  author    = {Rafailov, R. and Sharma, A. and Mitchell, E. and Manning, C. D. and Ermon, S. and Finn, C.},
  booktitle = {Advances in Neural Information Processing Systems (NeurIPS)},
  year      = {2023}
}

@article{tong2026scitaste,
  title   = {{AI} Can Learn Scientific Taste},
  author  = {Tong, Jingqi and Li, Mingzhe and Li, Hangcheng and Yang, Yongzhuo and Mou, Yurong and others},
  journal = {arXiv preprint arXiv:2603.14473},
  year    = {2026}
}

@article{zhu2026taste,
  title   = {{TASTE}: A Designer-Annotated Multi-Dimensional Preference Dataset for {AI}-Generated Graphic Design},
  author  = {Zhu, Haonan and Hirsch, Elad and Minetti, Alexandria and Nulty, Allison and Huffman, Ben and Mehta, Purvanshi},
  journal = {arXiv preprint arXiv:2605.20731},
  year    = {2026}
}

@article{pan2026tasteful,
  title   = {The Tasteful Agent: Measuring and Improving Taste in Long-Horizon Tasks},
  author  = {Pan, Wenbo and Liu, Zhichao and Liu, Shujie and Zeng, Jingying and Lin, Chin-Yew and Tang, Xianfeng and Lu, Yan and He, Qi and Jia, Xiaohua},
  journal = {arXiv preprint arXiv:2609.25804},
  year    = {2026}
}

\end{document}